\documentclass[preprint2]{aastex}

\newcommand{\kms}{\mbox{km\,s$^{-1}$}}

\slugcomment{accepted for publication in the Astronomical Journal}

\shorttitle{Pal5}
\shortauthors{Odenkirchen et al.}

\begin{document}

\title{Kinematic study of the disrupting globular cluster Palomar 5 
using VLT spectra\footnote{Based on observations obtained at the 
European Southern Observatory, Paranal, Chile (Program 67.D-0298A)}}

\author{
Michael Odenkirchen\altaffilmark{2}, 
Eva K. Grebel\altaffilmark{2}, 
Walter Dehnen\altaffilmark{2}, 
Hans-Walter Rix\altaffilmark{2},\\
Kyle M. Cudworth\altaffilmark{3}
}

\altaffiltext{2}{Max-Planck-Institut f\"ur Astronomie, K\"onigstuhl 17, 
D-69117 Heidelberg, Germany; odenkirchen@mpia-hd.mpg.de}
\altaffiltext{3}{Yerkes Observatory, University of Chicago, 373 West 
Geneva Street, Williams Bay, WI 53191; kmc@yerkes.uchicago.edu}

\begin{abstract}
Wide-field photometric data from the Sloan Digital Sky Survey have recently 
revealed that the Galactic globular cluster Palomar 5 is in the process of 
being tidally disrupted (Odenkirchen et al.\ 2001). 
Here we investigate the kinematics of this sparse remote star cluster using 
high resolution spectra from the Very Large Telescope (VLT). 
Twenty candidate cluster giants located within 6 arcmin of the cluster 
center have been observed with the UV-Visual Echelle Spectrograph (UVES) 
on VLT-UT2. The spectra provide radial velocities with a typical accuracy 
of 0.15~\kms. We find that the sample contains 17 certain cluster members 
with very coherent kinematics, two unrelated field dwarfs, and one giant 
with a deviant velocity, which is most likely a cluster binary showing fast 
orbital motion. 
From the confirmed members we determine the heliocentric velocity of the 
cluster as $-58.7\pm 0.2$~\kms. The total line-of-sight velocity 
dispersion of the cluster stars is $1.1\pm 0.2$~\kms\ (all members) or 
$0.9\pm 0.2$~\kms\ 
(stars on the red giant branch only). This is the lowest velocity dispersion 
that has so far been measured for a stellar system classified as a globular 
cluster. The shape of the velocity distribution suggests that there is 
a significant contribution from orbital motions of binaries and that the 
dynamical part of the velocity dispersion is therefore still substantially 
smaller than the total dispersion. 
Comparing the observations to the results of Monte Carlo simulations of 
binaries we find that the frequency of binaries in Pal\,5 is most likely 
between 24\% and 63\% and that the dynamical line-of-sight velocity 
dispersion of the cluster must be smaller than 0.7~\kms (90\% confidence 
upper limit). 
The most probable values of the dynamical dispersion lie in the range
$0.12 \le \sigma/\kms \le 0.41$ (68\% confidence). Pal\,5 thus turns out 
to be a dynamically very cold system. Our results are compatible with 
an equilibrium system. We find that the luminosity of the cluster implies 
a total mass of only 4.5 to $6.0 \times 10^3 M_\odot$. 
We further show that a dynamical line-of-sight velocity dispersion of 
0.32 to 0.37~\kms\ admits a King model that fits the observed surface 
density profile of Pal\,5 (with $W_0 = 2.9$ and $r_t = 16\farcm1$) and its 
mass.

\end{abstract}


\section{Introduction}
The globular cluster Palomar 5, an old halo cluster located at a distance 
of about 23~kpc from the Sun and 18.5~kpc from the Galactic center
(see Harris 1996), stands out through a number of unusual and 
extreme properties. These make it particularly interesting from the 
viewpoint of dynamics, cluster evolution, and galactic structure.  
First of all, Pal\,5 is extraordinarily sparse and faint. 
Its total luminosity of M$_V \simeq -5.0$ (Sandage \& Hartwick 1977, 
hereafter SH77) places it among the least luminous globular clusters 
that we currently know to exist in our Galaxy. 
Assuming a mass-to-light ratio typical of other globular clusters this 
luminosity corresponds to a stellar mass of about 
$1.3 \times 10^4$~M$_\odot$ (SH77) or $0.8 \times 10^4$~M$_\odot$ 
(Mandushev, Spassova \& Staneva 1991), which lies a factor of 30 below 
the median mass of Galactic globulars.     
Further, Pal\,5 has a very extended core and a low central concentration 
(see, e.g., Trager, King \& Djorgovski 1995). 
Finally, the faint part of the luminosity function of Pal\,5 is unusually 
flat (Smith et al.\ 1986, Grillmair \& Smith 2001), i.e.\ the fraction of 
low-mass stars in Pal\,5 is much smaller than in other galactic globulars. 
These peculiarities suggest that Pal\,5 has undergone strong dynamical 
evolution and mass loss and that it may be close to complete disruption. 
The hypothesis of ongoing dissolution was recently confirmed by the 
discovery of two massive tails of unbound former cluster members which 
spread from the cluster in opposite directions over an arc of several degrees 
(Odenkirchen et al.\ 2001).  
The stars observed in the tails make up at least 35\% of the total mass 
of cluster members and thus provide direct evidence for heavy mass loss 
from a strong tidal perturbation. 
Currently, Pal\,5 is therefore the best known example of a tidally 
disrupting globular cluster and is ideally suited for studying this 
phenomenon in-situ.

In order to reconstruct the mass loss history of Pal\,5 and to understand 
and interpret the density structures that are visible in its tidal tails 
one needs to simulate the cluster's dynamics with numerical methods.
The present-day velocity distribution in the cluster, in particular the 
velocity dispersion,
provides an important boundary condition for such simulations 
and is indispensable for deriving a realistic numerical model of the 
cluster's evolution in the tidal field of the Milky Way. 
However, the internal stellar velocities in Pal\,5 have not been measured 
to date. 
The only published measurements of radial velocities of stars in Pal\,5 are 
by Peterson (1985) and by Smith (1985). The spectral resolution 
of these measurements is sufficient to estimate the radial velocity of the 
cluster as a whole, but clearly too low to resolve the internal kinematics 
in the cluster. 
The results of Smith (1985) yield an upper limit of 4~\kms\ on the velocity 
dispersion in the cluster. 
Assuming that the cluster is in virial equilibrium the mass of 
$1.3 \times 10^4~M_{\odot}$ estimated by SH77 and a typical radius of 
about 20 pc imply a velocity dispersion of the order of 1~\kms.
On the other hand, since Pal\,5 has undergone tidal perturbations and is 
surrounded by a massive population of extratidal stars it would be conceivable 
that the above equilibrium model is not valid. 
Hence, any such prediction of the velocity dispersion in the cluster can only 
be a rough guideline.
The goal of this paper is to determine the internal kinematics of Pal\,5 
directly from observation. 
We thus set out to obtain high-resolution spectra for a number of cluster 
members to derive very precise radial velocities.  
Due to the rather large distance of the cluster, its deficiency in
bright giants, and the need for high spectral resolution, 
this project required an 8-m class telescope.

In Section 2 we provide details on the observations and the reduction
and analysis of the spectra. In Section 3 we then analyze the observed 
radial velocities, determine the cluster's velocity and the overall 
velocity dispersion along the line of sight, and compare the observed 
velocity distribution with a Gaussian model. In Section 4 we investigate 
the influence of binaries and derive constraints on the dynamical velocity 
dispersion of the cluster. In Section 5 we rederive the structural parameters 
and the total luminosity of the cluster from new photometric data and 
compare these parameters with the kinematics of the cluster in the framework 
of an equilibrium cluster model.     
In Section 6 we summarize and discuss our results.

\section{Observations and data reduction}
The stars that are used to probe the kinematics of Pal\,5 were selected with 
the help of multiband photometry from the Sloan Digital Sky Survey (SDSS; 
see Stoughton et al.\ 2002 and York et al.\ 2000 for overviews and 
Smith et al.\ 2002, Pier et al.\ 2002, Hogg et al.\ 2001, Gunn et al. 1998, 
Fukugita et al. 1996 for different technical aspects of the project).
We defined a sample of 20 program stars located within a radius of $6'$ 
around the center of Pal\,5, that have magnitudes in the range       
$15.0 \le i^* \le 17.7$, and appear as likely cluster giants according to 
their magnitude and colors (see Odenkirchen et al.\ 2001).
The target sample is presented in Table~1 and its properties are shown in 
Figures~1 and 2. 
These stars were observed with the echelle spectrograph UVES on ESO's 
Very Large Telescope (VLT). The observations were carried out in service 
mode during May and June 2001.
The spectra were taken in the instrument's red arm, using a dichroic beam 
splitter at 5800~{\AA} and a slit width of $1''$. 
They cover the wavelength ranges 4790~{\AA} $-$ 5760~{\AA} and 
5830~{\AA} $-$ 6810~{\AA} (recorded on two separate CCD chips) at a spectral 
resolution of about 40,000. 
Exposure times varied between 7 and 60 minutes and were chosen such that 
the spectra reach a signal-to-noise ratio of about 10. 
To enable precise wavelength calibration each observation of a star was 
followed by a ThAr lamp exposure taken immediately after the sky exposure. 
In addition to the program stars two radial velocity standards of type K1III 
(HD\,107328 and HD\,157457) from the list of Udry et al.\ (1999)  
were observed with the same instrument set-up in order to provide high 
quality template spectra and an absolute velocity reference.  

The reduction of the spectra (i.e., bias and background subtraction, 
flatfielding, order extraction, sky subtraction, wavelength calibration, 
rebinning, and order merging) was carried out with ESO's dedicated UVES 
reduction pipeline. 
The spectra were wavelength calibrated using the attached ThAr spectra 
and were rebinned to a linear wavelength scale of 0.030~{\AA}/pix (blue part)
and 0.035~{\AA}/pix (red part). Examples of the reduced and calibrated 
spectra are shown in Figure~3.

Radial velocities were determined by cross-correlating the spectra of 
the program stars with those of the two velocity standards. 
This was done using the routine {\sc fxcor} in the software package 
{\sc iraf}\footnote{
IRAF is distributed by the National Optical Astronomy Observatory}. 
We calculated separate cross-correlation functions in five distinct 
wavelength intervals, i.e., at
4910~{\AA} -- 5210~{\AA}, 5210~{\AA} -- 5460~{\AA}, 5460~{\AA} -- 5750~{\AA},
5850~{\AA} -- 6320~{\AA}, and 6320~{\AA} -- 6790~{\AA}. 
Hereby we obtained for each pair of a program star and a standard star five 
independent velocity measurements of nearly equal accuracy. 
The velocity shift between program star and template was determined by 
fitting a Lorentzian curve to the cross-correlation function in a range of 
$\pm 12.5$~\kms (i.e., 15 data points) around its highest peak. 
The results from the five wavelength intervals were averaged to define the 
relative velocity of each program star with respect to the standard star. 
The rms deviation of the individual results from the mean was used
to derive an empirical estimate of the random error of the velocity.  
Heliocentric corrections for the program stars and the templates, depending 
on their position and the time of their observation, were calculated with
the routine {\sc rvcorrect} in {\sc iraf}. 
Applying these corrections and adding the known absolute velocities of the 
template stars\footnote{According to Udry et al.\ (1999) the radial velocities
of HD\,107328 and HD\,157457 are +36.4~\kms\ and +17.8~\kms, respectively.}, 
the measured velocities were transformed to the heliocentric absolute system. 

The results of this procedure are presented in Table~2. The velocities 
are found to have accuracies between 0.05~\kms\ and 0.25~\kms. 
In most cases the random error is below 0.15~\kms. 
The results obtained with the two different velocity standards (see column 
(a) and (b) of Tab.\ 2) are in good agreeement with each other. 
There is a small mean offset of 0.14~\kms\ between the two sets of absolute 
velocities. When removing this offset, the remaining rms deviation is less 
than 0.03~\kms. 
Cross-correlation of the spectra of the two standard stars with one another 
yields velocities that differ from the nominal values by 0.14~\kms. The
difference agrees with the mean offset in the results for the program stars 
and shows that the zero-point of our absolute velocities has an uncertainty 
of this order.
We take for each program star the mean of the results obtained with either 
of the two standards as the best estimate of its absolute velocity.
Note, however, that the uncertainty of velocity differences is given by the 
individual random errors only. 

The last two columns of Table~2 give the velocities obtained 
by Smith (1985) and by Peterson (1985) for stars in common with those 
in our sample. The differences between our results and the previous ones 
are of the order of a few \kms.
It is easily seen that our results for different stars are in much closer 
agreement with each other than the previous measurements. 
This suggests that the differences between the previous results and ours 
are largely due to the lower precision of the previous measurements 
(and perhaps differences in the absolute calibration) and in general do 
not reflect true variations in the radial velocities of these stars. 
Variations from binaries are expected to occur mostly with smaller 
velocity amplitudes (see Sect.~4).

\section{Analysis of the kinematics}

\subsection{Cluster membership}

The spectroscopic observations show that most of the stars in our target 
sample are indeed giants and that many of them have almost identical 
velocities. 
Figure~4 shows the distribution of the observed velocities.
Seventeen of the twenty measured stars have radial velocities in the range 
between $-61.2$~\kms\ and $-56.9$~\kms. 
These are undoubtedly all members of the cluster.
Ten stars in this group stand out by having particularly coherent velocities 
that lie in an interval of only 1~\kms. This can be seen by the 
pronounced peak in Figure~4a and by the corresponding steep rise of 
the cumulative number counts shown in Figure~4b. 
Only three stars appear to be kinematically distinct from the cluster.
Their velocities deviate from the rest of the sample by about 6~\kms, 
14~\kms, and 35~\kms, respectively. 
Two of these stars turn out to be foreground dwarfs on the basis of a much 
larger width of their spectral lines (stars 4 and 12, see Fig.~3). 
They are definitely not belonging to the cluster and are discarded from 
the subsequent analysis.
The only doubtful case is star 15, which has the spectrum of a giant 
resembling those of other cluster members, but is set off from the cluster 
by about 14~\kms\ in radial velocity. Since this star is located at less 
than $2'$ angular distance from the center of the cluster it seems 
unlikely that it is an unrelated field giant at about the same spatial 
distance as the cluster and not a member of Pal\,5. 
From the surface density of field stars with magnitude and color similar
to star 15 (i.e., $\pm 0.15$~mag in magnitude and $\pm 0.05$~mag in color)
it follows that the expected number of such field stars within $2'$ from 
the cluster center is 0.1 while the expected number of cluster stars 
is 3.5.
Since we have observed three stars in this range of position, color, and 
magnitude the probability that (at least) one of them is a field star 
is $1-(3.5/3.6)^3 = 0.08$, only.

The cluster membership of all but stars 4 and 12 of our sample is 
independently confirmed in proper motion carried out by Cudworth, 
Schweitzer \& Majewski (in preparation). 
This study determines proper motions of about 500 stars in the 
field of Pal\,5 using microdensitometer scans of 25 plates from large 
reflectors ranging in epoch from 1949 to 1991. From the preliminary results 
of this work membership probabilities for our stars were derived 
in the way described in Dinescu et al.\ (1996). The probabilities are 
presented in Table 3. 
While stars 4 and 12 again prove to be non-members, the other 
stars have membership probabilities between 72\% and 99\% and thus 
qualify as cluster members. In particular, star 15 has a proper motion 
membership probability of 73\%, which is comparable to that of other stars 
with radial velocities close to the cluster mean, thus supporting its 
membership. 

The most plausible explanation then is that star 15 is a binary member 
of Pal\,5 and that the observed offset of its radial velocity (actually 
that of the primary component) is the result of temporal variations due 
to rapid orbital motion. For a binary in a circular orbit the orbital 
period $T$ is related to the orbital velocity $v_1$ of the primary by

\begin{eqnarray}
\frac{T}{\mathrm{yr}} = \left(\frac{30~\kms\ }{v_1} 
\frac{M_2}{M_1+M_2}\right)^3 \frac{M_1+M_2}{M_\odot} .
\end{eqnarray}

Assuming a mass of $0.8 M_\odot$ for the primary one finds that the 
orbital period of star 15 can at most be 2\,yr ($M_2=M_1$) and that 
it is likely to be of the order of a few months because the companion 
mass is probably smaller than that of the primary and because the projection 
on the line of sight has to be taken into account.
Therefore, even only one additional precise measurement of the radial 
velocity of this star should immediately reveal whether it is indeed a 
binary member of Pal\,5. 
For the present paper we take its occurence as an indication of the likely 
existence of binaries in Pal\,5, but omit it in the analysis of the 
velocities because of its large offset from the other cluster stars.

\subsection{Mean velocity and dispersion} 
The median of the velocities of the 17 certain members of Pal\,5
is $-$58.8~\kms, the arithmetic mean $-58.9\pm 0.3$~\kms. 
By successive omission of those stars that deviate most strongly from the 
median of the sample one obtains mean velocities between $-$58.6~\kms\ and 
$-$58.8~\kms. 
The particular subgroup of 10 stars whose velocities coincide within 
1~\kms\ has a mean velocity of $-58.7\pm 0.1$~\kms. 
Combining these results with the uncertainty of the zero-point of the 
absolute velocity scale we adopt the heliocentric velocity of the cluster 
$v_{r,cl} = -58.7\pm 0.2$. 
The rms dispersion of the individual velocities of the 17 cluster members 
with respect to this cluster mean is 1.14~\kms. The individual measurement 
errors are much smaller, and their contribution to this dispersion can 
be neglected. 
  
The colors and magnitudes of the stars reveal that the set of confirmed 
members consists of 13 normal red giants (RGB stars) and 4 asymptotic giant 
branch (AGB) stars (see Fig.~1). 
Comparing the velocities of the RGB and AGB stars we find that three of 
the four AGB stars have radial velocities that differ from the above 
cluster mean by more than 1~\kms. 
Among the RGB stars only three out of 13 have velocities beyond this limit. 
This suggests either that the two types of giants are somehow kinematically 
different or that the measurements of the AGB stars are affected by some 
kind of pulsational variations in their extended atmospheres. 
Spectroscopic studies of other globular clusters have found evidence for 
such `atmospheric jitter' among the most luminous red giants, i.e., close 
to the tip of the red giant branch (see, e.g., C\^{o}t\'e et al.\ 1996), 
but there is so far no information about a similar effect in less luminous 
AGB stars like those of our sample.
The presence of atmospheric jitter in our AGB stars could in principle be 
tested and eventually be removed through repeated measurements. 
However, since further observations of high precision are not available 
for these stars the question cannot be clarified at present.
If one excludes the AGB stars from the sample and considers only the 
subgroup of 13 RGB stars, the velocity dispersion drops from 1.14~\kms\ 
to 0.92~\kms.

\subsection{Dependence on magnitude and position}
In Figure~5 the individual radial velocities are plotted versus magnitude, 
angular distance $r$ from the cluster center 
($\alpha_c = 15^h16^m04.5^s$, $\delta_c = -00^\circ07'16''$, J2000.0) 
and position angle $\varphi$ (measured from north over east). 
Figure~5a shows that the mean velocity of the cluster members does not 
depend on the magnitude of the stars. 
Note that we also find no sign of a dependence of the measured velocities 
on the epoch of observation.
We thus can eliminate substantial systematic measuring errors as a function 
of brightness or epoch.   
Figure 5a also demonstrates that while there may be velocity jitter in 
the AGB stars there is clearly no sign for corresponding jitter in the 
brightest red giants of the sample since the velocities of these stars
are in extremely close agreement with each other.  

Figure~5b gives the impression that subgroups of stars at different angular 
distance from the center have very different velocity dispersions. 
This is partly, but not entirely due to the AGB stars, which are all 
located at $r > 2\farcm 5$ and hence differ from the spatial distribution 
of the normal giants. 
An F-test shows that the very low dispersion of $\sigma = 0.21$~\kms\ for 
the subgroup of the 4 innermost stars ($r \le 1\farcm 5$) is indeed
significantly smaller (99\% significance) than the dispersion for the 
remaining sample, even if the AGB stars are excluded. 
On the other hand, it turns out that the somewhat lower dispersion of the 
outermost stars (at $r > 4'$) as compared to the dispersion at medium 
distances from the center is not a statistically significant effect. 

In Figure~5c one finds weak indications that the observed velocities may 
contain azimuthal variations. Such variations could result from   
of a rotation of the cluster. However, there remains almost no evidence for 
such an effect if one leaves out the AGB stars.
Due to the limited number of data points the question of rotation cannot 
be investigated beyond the simple case of solid body rotation. 
By least-squares adjustment we find that if the velocities contain a solid 
body rotation component the angular velocity of the rotation and the 
position angle of the rotation axis would be 
$\omega = 0.25\pm 0.12$~\kms\,arcmin$^{-1}$ and $PA = +15\pm 31^\circ$, 
respectively.
When subtracting this hypothetical rotation from the observed velocities 
the velocity dispersion of the n=17 sample reduces only slightly (from 
1.14 to 1.08~\kms). For the subsample without AGB stars we likewise obtain 
$\omega =0.17\pm 0.16$~\kms\,arcmin$^{-1}$, i.e., the rotation velocity 
is at the border of statistical significance, and the rotation model yields
no reduction of the velocity dispersion.  
Therefore we conclude that rotation is not clearly detectable and that it 
does not provide an important contribution to the kinematics of the cluster.  
This conclusion also holds for the possibility of an expansion of the 
cluster along a preferred spatial direction since from the observational 
point of view such an effect is equivalent to a rotation.

\subsection{Comparison with an isothermal system} 

If the cluster maintains a state of dynamical quasi-equilibrium one expects 
the kinematics in its inner region to be isothermal, which means that the 
line-of-sight velocity distribution of objects with approximately equal
masses must (in the absence of other effects) be Gaussian. 
Many stellar systems do indeed show velocity distributions that are 
approximately Gaussian. 
However, in the case of a cluster with low velocity dispersion it may happen
that contributions from atmospherically induced variations and from orbital 
motion of binaries are of non-negligible size, and that the observed 
line-of-sight velocities are hence not completely dominated by the dynamics 
of the cluster. Concerning contributions from binaries, it may also be 
relevant that a small velocity dispersion in the cluster is favorable to 
the survival of binaries during the cluster's dynamical evolution.  
The impact of velocity anomalies in the AGB stars, possibly caused by 
atmospheric jitter, was discussed in Section~3.3. 
In order to find out if binaries play a significant role in Pal\,5 we 
compare the observed velocities to a simple Gaussian representing an 
isothermal system of single stars. 

The Gaussian that best fits the observations can be found by the method 
of maximum likelihood.
When observational errors are neglected, the dispersion parameter $\sigma$ 
of the best-fit Gaussian is simply given by the rms dispersion of the 
observed velocities. 
In the general case with observations $v_{r_i}$ and errors $\epsilon_i$, one 
calculates the probabilities $p_i$ of the observations by convolving the 
Gaussian model $\Phi_d$ with the error distributions $\Phi_{\epsilon_i}$, 
i.e.\ $p_i = (\Phi_d \ast \Phi_{\epsilon_i})(v_{r_i})$, and maximizes the 
total probability (or likelihood) $\mathcal{L} = \prod_i p_i$ as a function 
of the parameter $\sigma$. 
For the complete sample of $n=17$ stars the solution is $\sigma = 1.14$~\kms\ 
with uncertainties of $+0.24$~\kms\ and $-0.18$~\kms.
For the $n=13$ subsample of RGB stars we obtained $\sigma = 0.91$~\kms\ 
with uncertainties of $+0.23$~\kms\ and $-0.18$~\kms. 
The given uncertainties describe the interval around the maximum of 
$\mathcal{L}$ that contains 68.3\% of the likelihood integral $\int 
\mathcal{L} d\sigma$.

To test the agreement between the maximum likelihood Gaussian model and the 
observations we generated a large number of artificial samples of $n$ 
velocities based on the best-fit $\sigma$ and the observational errors, 
and compared the likelihood of the observed velocities with the likelihood 
of the simulated velocities. 
The experiment reveals that the likelihood values of the observed samples 
fall close to the median of the likelihood distribution of the corresponding
simulated samples, i.e., 47\% ($n=17$ case) or 45\% ($n=13$ case) of the 
simulated samples have likelihood values that are smaller than the 
likelihoods of the observed samples. This means that from the viewpoint of  
likelihood statistics the best-fit Gaussian is an acceptable model for the 
observed velocities. It would however be wrong to conclude that a purely 
Gaussian model provides an optimal description of the data.
 
In Figure~6 the data and the best-fit Gaussian models are compared by 
plotting their cumulative distributions. Here it is seen that the agreement 
between the data and the models is in fact less than satisfactory.
For both samples there are clear systematic differences between the data and 
the model. The observational data are characterized by a steep rise at small 
velocities and a flat tail extending from 0.3 to 2.4~\kms. 
A Gaussian distribution is unable to approximate this shape in an 
appropriate way. 
In order to maintain the idea of isothermal cluster kinematics we thus need 
to assume the existence of an additional velocity component.   

This suggests that there are indeed significant contributions from orbital 
motion of binaries. These binaries may either be primordial (as in young 
open clusters and the field) or may have formed by close stellar encounters 
during the evolution of the cluster.
Recent searches for binaries in other galactic globulars have revealed 
that globular clusters are not generally deficient in binaries as compared 
to the local population of field stars and that clusters with low central 
density and/or indications of strong tidal mass loss tend to have enhanced 
binary frequencies (see McMillan, Pryor \& Phinney (1998) and references 
therein).   
Moreover, star 15 provides a direct hint that the existence of binaries 
in our sample is likely (see Section 3.1).  
If this object is indeed a binary member of Pal\,5 with a relatively short 
period of a few months as assumed in Section~3.1, then it is natural to 
expect that the sample also contains binaries with longer orbital periods 
of up to several decades. Such systems will typically produce shifts of 
the order of 1~\kms\ in the observed velocities. 
This can be seen by considering the simple but typical example of a binary 
with a total mass of 1~$M_\odot$ and a mass ratio of 1:4 
(i.e., $M_1=0.8 M_\odot, M_2=0.2 M_\odot$). Assuming a circular orbit one 
finds that a radial velocity of the primary of about 1~\kms\ along an 
average line of sight is obtained with an orbital period of about 30 years. 

The presence of significant contributions from binaries of course makes 
the determination of the dynamical velocity dispersion of the cluster 
more difficult.
The above example shows that a direct proof of the binary nature of 
individual stars and the determination of center-of-mass velocities 
through repeated velocity measurements is not categorically impossible, 
but would require a large amount of observing time over a long time scale 
(except for star 15). 
Another way (and currently the only practicable one) to eliminate the  
influence of orbital motion and hence to determine the dynamical velocity 
dispersion is by estimating the frequency of binaries and their contribution 
to the observed velocity dispersion in a statistical manner. 
This approach works by incorporating the effects of orbital motion into 
the model for the distribution of the observed radial velocities,
as will be described in the next section.

\section{Simulation of binaries}

Since the binarity of the observed objects and their orbital motions are
not individually known one may assume that each object has a probability $x_b$
of being a binary and consider a variety of possible systems with 
statistically distributed orbital parameters. If the distribution functions 
of the orbital parameters of the binary population are known, then the 
statistical impact of their orbital motions on the radial velocity of the 
primaries and the resulting distribution of observable velocities can be 
determined by a Monte Carlo simulation. 
For binaries in globular clusters one faces the problem that specific 
information on the distributions of their parameters is lacking. 
Hence one must either work with plausible general assumptions or refer 
to the empirical parameter distributions for field binaries in the local 
neighborhood, hoping that these are not prohibitively wrong for cluster
binaries.
We decided to try both ways and thus generated different sets of artificial 
binary populations in the following way:
   
The mass of the primary component was chosen as $M_1 = 0.8~M_\odot$ for 
normal red giants and $M_1 = 0.5~M_\odot$ for AGB stars. 
The mass $M_2$ of the secondary was modelled by a log-normal mass 
function in the mass range $0.1\,M_\odot \le M_2 \le M_1$, i.e.\ a 
Gaussian for $\log M$, with mean $<\log (M/M_\odot)> =  -0.38$ and dispersion 
$\sigma_{\log (M/M_\odot)} = 0.35$. 
This mass function closely resembles the empirical mass distribution of 
secondaries around nearby solar-type stars found by Duquennoy \& Mayor 
(1991). 
Alternatively, we also tried simpler choices as, for example, $M_2 = M_1$, 
$M_2 = 0.4 M_\odot$, or $M_2 = 0.2 M_\odot$.   
For the semi-major axis $a$ of the binary orbits we adopted a Gaussian 
in $\log (a/\mathrm{AU})$  with mean $<\log (a/\mathrm{AU})> = 1.5$ and 
dispersion $\sigma_{\log (a/\mathrm{AU})} = 1.5$. 
This model is in agreement with empirical semi-major axis distributions 
as, e.g., given by Heintz (1969), and with the period statistics of local 
G and K dwarfs given by Mayor et al.\ (1992). We note that this 
distribution was used in truncated form (see below). 
For the orbital eccentricity $e$ we used as standard case a flat 
distribution $f(e)=const$ in the range $0.1 \le e \le 0.8$ and $e=0$ 
for periods $P < 0.3$~yr, taking into account observational results 
by Duquennoy \& Mayor (1991), Mayor et al. (1992) and Latham et al. (1998).
In addition, we also ran simulations with either $e = 0$ (circular orbits) 
or $f(e)=2e$ (so-called thermal orbits) as extreme alternatives.

In a cluster environment wide binaries are unlikely to survive stellar 
encounters if their binding energy is less than the typical kinetic 
energy of relative motion between cluster members. 
Using a velocity dispersion of 2 to 4~\kms\ (as an estimate for earlier 
evolutionary stages of Pal~\,5) the formula given in equation~(1) of 
Pryor et al.\ (1996) yields upper limits for the semi-major axis of 
100~AU to 25~AU, respectively. 
We thus tentatively truncated the distribution of $a$ at different values 
in this range and after some testing set the upper limit of $a$ to 50~AU.  
At the lower end the range of distances between the two binary 
components also has to be truncated because the primaries in our sample 
are giants and thus have large radii which can surpass the Roche limit 
for mass transfer. 
Roche overflow would increase the separation between the components 
and decrease the brightness of the system (see, e.g., the discussion by 
Pryor et al.\ 1988). 
Therefore we calculated approximate radii for our giants using the 
radius-magnitude relation shown in Figure~6 of C\^{o}t\'e et al.\ (1996), 
determined the distance of the inner Lagrange point L1 from the primary 
component at pericenter, and neglected all cases in which this distance was 
smaller than the estimated stellar radius ($10 \le R/R_\odot \le 40$). 
Note that this constraint precludes the existence of very hard binaries 
in our sample.

From each simulated set of binaries we derived a characteristic radial 
velocity distribution $\Phi_b$ by projecting the orbital velocities of the 
primary components onto isotropically distributed line-of-sight directions 
at randomly chosen fractions of the orbital period.  
Examples of such distributions are shown in Figure 7. These distributions 
have long high-velocity tails which distinguish them from a Gaussian.
To describe the radial velocities of a sample of binaries in a cluster 
with isothermal kinematics the binary radial velocity distribution $\Phi_b$ 
needs to be convolved with a Gaussian $\Phi_d$ that models the dynamical 
velocity dispersion as in Section~3.4. 
Furthermore, for a cluster with a fraction $x_b$ of binaries the velocity 
distribution must be a composite of $\Phi_d \ast \Phi_b$ and $\Phi_d$, namely

\begin{eqnarray}
\Phi = x_b \cdot (\Phi_d \ast \Phi_b) + (1-x_b) \cdot \Phi_d\ . 
\end{eqnarray}
     
This yields an extended model $\Phi$ with adjustable parameters $\sigma$ 
and $x_b$ that should allow a better match with the observed velocities 
provided that the distributions of the orbital parameters of the binaries 
have been set appropriately.  
The model was fitted to the observations in the same way as in Section 3.4, 
namely by convolving it with the observational errors, calculating the 
individual probabilities $p_i$ of the observed velocities, and then 
maximising the probability product $\mathcal{L}$ (or actually its 
logarithm $\log \mathcal{L} = \sum_i \log p_i$) as a function of $\sigma$ 
and $x_b$. 
We calculated maximum likelihood solutions for a variety of choices of 
the binary parameters $M_2$ and $e$. 
The results are presented in Table~4. The upper part of the table refers 
to the entire sample and the lower part to the subsample of red giants. 
The quoted uncertainties in $\sigma$ and $x_b$ describe the parameter 
range that comprises 68.3\% of the integrated likelihood. 
In addition, we provide the value of the 90\% confidence upper limit of 
$\sigma$, which is denoted $\sigma_{90\%}$. 
For comparison, the first line in each part of Table~4 repeats the single 
star solution from Section 3.4. 

It turns out that most of the binary models indeed enable solutions with 
higher maximum likelihood than the single star model. 
The only exception is the extreme case of equal mass binary components  
($M_1=M_2$), which fails to produce an improved fit to the velocity  
distribution of the $n=17$ sample and thus proves to be inadequate. 
Comparing the values of $\ln \mathcal{L}$ in Table~4 it is seen that  
the improvements achieved with the different binary models are generally 
larger for the $n=13$ subsample than for the complete $n=17$ sample. 
This supports the assumption that the increased velocity dispersion of the 
AGB stars has a different origin (see Sect.~3.2) and is not due to binarity. 
The best fit to the observed velocities is obtained with the model that 
is based on low-mass companions ($M_2=0.2 M_{\odot}$) and preferentially 
high orbital eccentricities ($f(e)=2e$). This holds for both samples.
Our so-called standard case which is closest to the properties of the 
local field binaries is found to be less adequate. 
Other cases with more massive secondary components or low eccentricities 
are also seen to be less in agreement with the observed velocities.

Although the solutions are of different quality we find that the results
for the fit parameters $\sigma$ and $x_b$ do not depend much on the 
details of the models. For the remainder of this section we focus on the 
results for the $n=13$ subsample of red giants.
Table~4 shows that all models require best-fit binary fractions between 
0.30 and 0.42, with uncertainties of about $\pm 0.15$ to $\pm 0.20$. 
The best-fit values of the dispersion parameter $\sigma$ fall in all cases 
close to 0.22~\kms, and the probable values lie between 0.12~\kms\ to 
0.44~\kms. Hence it is likely that the dynamical velocity dispersion 
is by at least a factor of two lower than the directly measured total 
velocity dispersion. 
Figure~8 shows the likelihood distribution in the plane of the fit 
parameters $x_b$ and $\sigma$ for the model that yields the best fit to 
the observations (i.e., case (h) of Tab.~4). 
It is seen that there exists a unique likelihood maximum at low $\sigma$
and intermediate $x_b$. However, this maximum is relatively broad and 
has a tail towards higher $\sigma$, so that the resulting ranges of probable 
values (68.3\% confidence) extend from 0.24 to 0.63 in $x_b$ 
and from 0.12 to 0.41~\kms\ in $\sigma$. 
Higher values of $\sigma$ cannot be entirely excluded but are less likely. 
To set a reasonable upper limit one can state that with 90\% confidence 
the dynamical velocity dispersion of the cluster is not larger than 0.7~\kms.  

In Figure~9, we compare the cumulative distribution of observed velocities 
with the corresponding predictions by two of our models with binaries, namely 
the standard model and the one that yields the best fits. Other than the single 
Gaussians shown in the analogous Figure~6, these models can (due to the 
inclusion of binaries) approximate both the steep inner rise and the extended 
tail of the observed velocity distribution. The standard model however is 
seen to be somewhat less adequate than the other model since it predicts too 
many stars with $|\Delta v_r| > 1.5$~\kms\ in comparison to the observations.  
    
\section{Velocity dispersion versus structural parameters, 
luminosity, and mass-to-light ratio}

A key question in connection with the determination of the 
velocity dispersion is:
How does the result compare to other fundamental parameters of the 
cluster? We combine the discussion of this question with a 
redetermination of Pal\,5's size, structure, luminosity, and mass using 
the SDSS data and the HST luminosity function for main-sequence stars 
of Pal\,5 by Grillmair \& Smith (2001).

\subsection{Surface density profile and best-fit King model}

The wide-field photometry from SDSS provides us with the possibility
to derive an improved radial density profile for Pal\,5. Using the advantage 
of selective star counts the cluster's profile can be traced down to five 
times lower surface density than in former work by Trager et al.\ (1995).
We separated cluster members from field stars through an efficient 
color-magnitude filter (see Odenkirchen et al.\ 2001) and determined 
the surface density of the members by counts in circular annuli out to a 
radius of 100 arcminutes.
In the outer annuli the member counts are in some places influenced by an 
overlap with the tidal tails.  
This was taken into account by splitting the field into four $90^\circ$ 
sectors pointing towards north, south, east, and west.  
The profiles obtained for the northern and southern sector and for the 
eastern an western sector are compared in Figure 10.
Since the cluster overlaps with the tails only in the northern and southern 
sector, the counts obtained in the eastern and western sector yield a profile 
that is free of contamination by tidal tail stars and describes the true size 
and structure of the cluster. 

Under the assumption of dynamical quasi-equilibrium the density profile of the 
cluster should match the profile of a so-called King model, i.e., a truncated 
isothermal sphere (King 1966). 
We thus took the 'clean' surface densities measured in the eastern 
and western sector of the field, fitted a series of King profiles by means 
of weighted least squares, and determined the best-fitting King (1966) model 
for the cluster. 
The best-fit model has $W_0 = 2.9$, which is equivalent to $c = 0.66$ for 
the concentration parameter, and a limiting radius $r_t = 16\farcm 1\pm 
0\farcm 8$. The corresponding core radius is $r_c = 3\farcm6\pm 0\farcm2$. 
At the assumed distance of $d=23.2$~kpc of Pal\,5 this means a linear core 
radius of 24.0~pc. 
Figure 10 shows the best-fit King model (plus constant foreground density) 
by a solid curve. The agreement between the measurements and the model is 
satisfactory beyond the core radius, but the model fails to describe the 
constant surface density in the central $3'$ of Pal\,5 and the subsequent 
abrupt decline. This may be an indication that the dynamical state of 
the cluster is in fact somewhat different from a King model. Nevertheless, 
the best-fit King model remains a useful approximation because it provides a 
relation between mass and velocity dispersion.

The mass of a truncated isothermal sphere with spatial 
parameters $r_c$ and $c$ and with a line-of-sight velocity dispersion 
$\sigma_{los}$ is given by

\begin{eqnarray}
M = \frac{9}{4{\pi}G}\ r_c\ \mu\ (\beta\,\sigma_{los})^2
\end{eqnarray}

\noindent (see King 1966). Here, $\mu$ denotes a normalized mass parameter  
that depends on $c$ and that has to be determined by numerical integration. 
For a model with $W_0=2.9$ one finds $\mu=4.9$. The correction factor $\beta$ 
depends on the model and on the area within which the velocity dispersion is 
sampled. 
If $\sigma_{los}$ is determined as an average line-of-sight dispersion
over a region of one core radius, as it is the case with our observations, 
the appropriate correction factor $\beta$ for a model with $W_0 = 2.9$ 
is $\beta = 1.5$ (see Binney \& Tremaine 1987, p.236).    
Inserting these values into equation (3) we obtain the relation: 

\begin{eqnarray}
\frac{M_{Pal5}}{M_\odot} = 4.4 \cdot 10^4 \left(\frac{\sigma_{los}}{\kms}\right)^2  
\end{eqnarray}
 
In order to compare this with the observed velocity dispersion, one needs  
an independent estimate of the cluster's mass from its luminosity. 

\subsection{Total luminosity and mass } 

To determine the total absolute $V$ band magnitude of the cluster we 
integrated the flux of all SDSS stars that lie in the region of the 
giant and subgiant branch, horizontal branch and main-sequence of 
Pal\,5 and within the limiting radius $r_t=16'$ of the cluster. 
The integration was carried out down to a magnitude limit of $V=21.75$. 
Visual magnitudes were derived from SDSS $g^*$ and $r^*$ magnitudes using the 
transformation $V = 0.4 g^* + 0.6 r^* + 0.23$. This relation was set up 
by comparing SDSS photometry and CCD photometry in $V$ from Smith et 
al.\ (1986) for 18 stars in the color range $-0.2 \le g^*-r^* \le 1.4$.
We corrected the integrated flux for the contribution of residual 
field stars by subtracting the statistically expected luminosity of 
field stars as estimated in nearby fields. 
The resulting integrated magnitude of Pal\,5 from stars down to the 
limit of $V = 21.75$ is $m_V = 12.24\pm0.07$. 
Using a distance modulus of $16.83\pm 0.20$ ($d = 23.2\pm 2.3$~kpc) 
the absolute magnitude of the cluster down to this limit is 
$M_V = -4.59\pm 0.20$.  

The missing flux of cluster stars fainter than $V = 21.75$ was 
estimated using the deep luminosity function (LF) of Grillmair \& Smith 
(2001). 
We rescaled this LF to the SDSS star counts in the range 
$19.75 \le V < 21.75$ and then integrated the 
predicted flux from $V=21.75$ down to $V=27$, where the LF is flat 
and the contribution to the total flux becomes negligible. 
This yields an additional contribution to the integrated magnitude of the 
cluster of 0.18~mag. Our result for the total absolute magnitude of Pal\,5 
is thus $M_{V} = -4.77\pm 0.20$. 
This is marginally lower than the estimate of $M_{V} = -5.0$ obtained 
by SH77 and corresponds to a total V band luminosity of 
$(L/L_\odot)_V = 7.2 \times 10^3$.  (using $M_{V} = 4.87$ for the Sun).  

One way of deriving the mass of the cluster from its luminosity is by 
assuming that the mass-to-light ratio of Pal\,5 is not substantially 
different from those of other Milky Way globular clusters. 
According to the empirical mass-luminosity relation by Mandushev, Spassova 
\& Stanova (1991) which reads 

\begin{eqnarray}
\log(M/M_\odot) = -0.456\,M_V + 1.64
\end{eqnarray}

the mean mass-to-light ratio of galactic globulars varies between 
$M/L_V = 1.1$ for faint clusters ($M_V = -6$) and $M/L_V = 1.8$ for 
very bright clusters ($M_V = -10$).
The estimates of the masses and mass-to-light ratios of individual 
clusters deviate from this relation by at most a factor of two. 
Since the relation is calibrated with clusters that have $M_V \le -5.6$
its application to Pal\,5 involves an extrapolation towards fainter 
magnitudes. Equation (5) then provides an estimate of the cluster's mass 
of $M/M_\odot = (6.5\pm 1.5) \times 10^3$, corresponding to 
$M/L_V = 0.90\pm 0.20$.  
For two reasons this estimate may not appear completely convincing by itself. 
First, the extrapolation to $M_V < -5.6$ is uncertain. Second, it is not 
a priori clear that Pal\,5 fits into the above mean relation for other 
globular cluster because its main-sequence luminosity function is known 
to be flatter than that of other clusters.

Therefore we derived the mass of the cluster also in a more direct 
way using the luminosity function of the cluster and theoretical stellar 
masses from a 14~Gyr isochrone by Bergbusch \& Vandenberg (1992).  
Same as for the calculation of the total luminosity the luminosity function 
of Grillmair \& Smith (2001) was rescaled to the SDSS counts for the 
entire cluster in the range $19.75 \le V < 21.75$. 
We then multiplied in each interval of width 0.5~mag from $V = 14.75$ down 
to $V = 28.75$ the number of stars with the stellar mass for the mean 
absolute magnitude of the interval and computed the sum of these masses, 
i.e., the integral of the mass function. 
For the faintest three bins from $V=27.25$ to $V=28.75$ for which 
no star counts are available from observation we assumed that the luminosity 
function at $V \ge 27$ is constant. This allowed us to extend the 
integral of the mass function down to $0.17 M_\odot$ and hence to be complete 
down to the transition from stars to brown dwarfs. The mass contribution 
from the faintest three bins is of the order of 10\%.
The total mass of the cluster according to this second approach is 
$(5.2\pm 0.7)\times 10^3 M_\odot$, which corresponds to 
$M/L_V = 0.73\pm 0.10$.
The quoted uncertainty has been estimated by considering the changes that 
occur when varying the age of the isochrone, the radius of the field 
encircling the cluster, the transformation between SDSS and standard 
magnitudes, and the distance modulus of the cluster.    
The result shows that the extrapolation of the mean mass-luminosity relation 
in equation (5) down to the absolute magnitude of Pal\,5 is principally 
correct, but that this may still slightly overestimate the mass of the 
cluster.
We conclude that the mass of Pal\,5 as revealed by its luminosity 
is in the range $4.5 \times 10^3 \le M_{Pal5}/M_\odot \le 6.0 \times 10^3$, 
which corresponds to $0.63 \le M/L_V \le 0.83$ for the mass-to-light ratio. 

\subsection{Implications for the velocity dispersion} 
By putting the empirical mass limits from the end of the previous section 
into equation (4) we obtain the following result:
A King model that fits the stellar surface density of Pal\,5 and has 
the appropriate mass of 4.5 to $6.0 \times 10^3 M_\odot$ 
requires an observable line-of-sight velocity dispersion between 
$\sigma_{los}=0.32$~\kms and $\sigma_{los} = 0.37$~\kms. 
Velocity dispersions at the level of 0.7~\kms\ or higher are not in 
agreement with an appropriate King model because one would need to
assume a high mass-to-light ratio of $M/L_V \ge 3$, which from
the results of Section 5.2 is unrealistic.     
For $\sigma_{los} = 0.91$~\kms\ equation (4) leads to a predicted 
equilibrium cluster mass of $3.6 \times 10^4 M_\odot$. 
This shows that if one denies the presence of binaries the resulting 
velocity dispersion is much too high to be consistent with the mass 
that is revealed by the cluster's  luminosity. 
However, when taking the binaries into account as done in Section 4, 
the estimates of the line-of-sight velocity dispersion of the cluster
decrease in such a way that a consistent equilibrium model with normal M/L 
is then viable and a velocity dispersion in excess of the equilibrium 
value is unlikely. 
More specifically, the line-of-sight velocity dispersion must lie in the 
upper third of 
the 68\% confidence interval of probable values of $\sigma$ in order to 
enable an equilibrium model that is in agreement with the surface density 
profile and the mass of the cluster.
Dispersion values near the point of the highest likelihood (i.e., at 
$\sigma \approx 0.22$~\kms) on the other hand are too low to be consistent 
with an equilibrium model of Pal\,5 since this would require a very small 
mass-to-light ratio of $M/L_V \approx 0.3$, which again contradicts the 
results of Section 5.2.  
  
\section{Summary and discussion}

Our study has shown that the radial velocities of the giants in the 
cluster Pal\,5 are tightly concentrated around a mean velocity of 
$-58.7$~\kms\ and have an overall dispersion of at most 1.1~\kms. 
Only one out of a total of 18 observed giants appears as a kinematic outlier. 
Statistical arguments suggest that this outlier is also a member of the 
cluster. Its velocity offset of about 14~\kms\ with respect to the other 
cluster members leads to the conclusion that it is most likely a binary 
with an orbital period of a few months. This makes it a very interesting 
case because the occurrence of a binary with a relatively short period 
close to the center of Pal\,5 would fit with the general idea that such 
binaries, if not primordial, are formed by stellar encounters 
in globular cluster cores during the dynamical evolution the cluster.
(see, e.g., Meylan \& Heggie 1997). Such binaries are believed to be 
an important energy source that supports the evaporation of the cluster. 
However, the assumption that the above object is such a rapidly orbiting 
binary still needs to be confirmed. 

A peculiar feature in our velocity data is that the few AGB stars in the 
sample show a significantly higher velocity dispersion than the stars on 
the RGB. In principle, such a difference could result from the lower 
mass of the AGB stars. In practice, however, the mass ratio of about 
0.5:0.8 between AGB and RGB stars is by far not small enough to explain 
the observed kinematic differences. The possibility that binarity causes 
the enhanced velocity dispersion of the AGB stars is also unlikely because 
there is no reason to believe that AGB stars should have a much higher 
frequency of binaries than RGB stars. 
We thus tend to believe that the enhanced velocity dispersion among 
the AGB stars is due to so-called atmospheric jitter, i.e., due to 
pulsations in the extended atmospheres of these evolved stars. 
This assumption is motivated by the fact that very luminous RGB and AGB 
stars in other globular clusters and in the field often exhibit 
photometric and spectroscopic variations due to pulsating atmospheres. 
A possible caveat however is that the stars in our sample do not belong 
to those most luminous types of giants since they are more than 1~mag  
fainter than the tip of the red giant branch. Moreover, the 
RGB stars in our sample that have the same brightness as the AGB stars 
show a very small velocity dispersion that leaves no room for any jitter.   
It will therefore be necessary to check the hypothesis of atmospheric 
pulsation by collecting further observations of these stars.     
As long as the nature of the enhanced velocity dispersion of the AGB stars 
is not clear, one must conclude that these stars are not useful for 
investigating the dynamics of the cluster. 

Without the AGB stars (and the above outlier of course) the observations 
yield an overall line-of-sight velocity dispersion of 0.9~\kms. 
This value sets a strict upper limit on the dynamical line-of-sight 
velocity dispersion of the cluster. However, by further analysis of the 
data we saw that this limit overestimates the dynamical velocity 
dispersion substantially because orbital motions of (long-period) binaries 
have an important influence that cannot be neglected. 
The fact that the distribution of the binary-induced velocities is 
different from the Gaussian distribution for an isothermal cluster  
allowed us to distinguish the two velocity components and to estimate 
the fraction of binaries $x_b$ and the dynamical line-of-sight velocity 
dispersion $\sigma$.
Using Monte Carlo simulations we constructed synthetic velocity 
distributions for different types of binary populations, combined them 
with the Gaussian model of the cluster kinematics, and fitted 
these composite models to the observed velocity distribution. 
The results of this procedure suggest that the binaries in our sample 
have low mass companions of about $0.2 M_\odot$ and orbits of very high 
eccentricity.
Interestingly, a binary population with the orbital characteristics of 
local field binaries is less in agreement with the observed velocity 
distribution. 
This might be an indication for differences between cluster binaries 
and field binaries that arise from the special cluster environment and 
the dynamical evolution in the cluster.  
On the other hand, the details of the different binary models turned 
out to be rather unimportant for the determination of the parameters 
$x_b$ and $\sigma$ since all models with a substantial fraction of 
secondary masses $\le 0.4 M_\odot$ led to similar estimates.  
We thus arrived at the conclusion that with 68\% confidence the frequency
of binaries in our sample of giants is in the range from 0.24 to 0.63,
and that with the same confidence level the dynamical velocity dispersion of 
the cluster lies between 0.12~\kms\ and 0.41~\kms. The latter means that 
the dynamical line-of-sight velocity dispersion is by at least a factor 
of two smaller than the overall velocity dispersion and that the contribution 
from binaries accounts for more than 50\% of the overall dispersion.

Are these results credible? 
The binary frequency of roughly $40\%\pm 20\%$ is consistent with 
results from systematic searches for spectroscopic binaries in other 
galactic globular clusters. For normal clusters such studies have 
reported binary frequencies of about 5\% to 15\% per decade of period 
or estimates of overall binary frequencies of about 20\% to 30\% 
(see the reviews by Meylan \& Heggie (1997) and by McMillan, 
Pryor \& Phinney (1998)).    
Moreover, there is evidence that clusters with low density and/or strong 
tidal mass loss can have two to three times higher binary frequencies 
than normal clusters (see, e.g., Pryor, Schommer \& Olszewski (1991) 
and Yan \& Cohen (1996)).
If we take $x_b=0.24$ from the low end of our 68\% confidence interval 
as a conservative estimate, this corresponds to three binaries in the 
sample of 13 RGB stars. Assuming that at least one of these binaries, 
but perhaps two or even all three of them are responsible for the most
deviant velocities in the sample, we can successively omit those three 
stars that have the highest absolute velocities with respect to the cluster 
mean and calculate the velocity dispersion of the remaining subsample.
Hereby we obtain velocity dispersions of 0.64~\kms, 0.51~\kms, and 
0.32~\kms\ (corrected for measurement errors), respectively.
This simple test illustrates that even modest assumptions on the frequency 
and influence of binaries result in a substantial reduction of the velocity 
dispersion and easily bring it down to the level of 0.5~\kms. 
By successive omission of the three most deviant velocities
the velocity distribution also becomes progressively 
more consistent with an isothermal distribution. 
The differences between the empirical distribution and the corresponding 
Gaussian model then have Kolmogorov-Smirnov significance levels  
of 48\%, 15\% and 10\%, respectively, while the significance of the 
differences is 78\% for the original sample. 

Another interesting question is: Does the observed velocity dispersion 
admit an equilibrium cluster model that is consistent with other 
fundamental parameters of Pal\,5? 
To investigate this point we derived the surface density profile and 
the total luminosity of the cluster from recent photometric data and 
determined the King model that best fits the density profile.  
The parameters of the best-fit King profile are $W_0=2.9$, $r_t=16\farcm1$. 
According to its absolute magnitude, which we found to be 
$M_V = -4.77\pm 0.20$, and by extrapolation with a general globular 
cluster mass-luminosity relation Pal\,5 has a mass in the range 
$5 \times 10^3 M_\odot$ to $8 \times 10^3 M_\odot$, equivalent to 
a mass-to-light ratio between 0.7 and 1.1. 
A direct determination of the mass using the cluster's luminosity function 
and theoretical stellar masses along a 14~Gyr isochrone gave a similar  
result, namely a total mass between $4.5 \times 10^3 M_\odot$ and 
$6.0 \times 10^3 M_\odot$. We showed that
the best-fitting King model comes into agreement with this mass if the 
line-of-sight velocity dispersion is between 0.32 and 0.39~\kms. 
Therefore, a line-of-sight velocity dispersion that lies in the upper 
third of our 68\% confidence interval for $\sigma$ does indeed admit a 
consistent equilibrium model for Pal\,5.
 
By testing a variety of binary models we saw that the range of probable 
values of $\sigma$ does not critically depend on the choice of the 
orbital parameters of the binaries. In order to arrive at probable   
velocity dispersions that significantly exceed the equilibrium dispersion 
of the cluster one would thus need to assume that the cluster has only a 
small fraction of binaries. 
This case cannot be strictly ruled out but it appears unlikely and is not 
supported by the solutions obtained in Section 4. 
Based on the more plausible alternative assumption that binaries are not 
rare in Pal~5 we draw the conclusion that although the cluster has obviously 
experienced strong tidal perturbation and heavy mass loss, the remaining 
central body still presents a bound stellar system that is close to a state 
of dynamical quasi-equilibrium. 

This conclusion is in agreement with results that we have recently  
obtained by simulating the dynamics of Pal~5 in the tidal field of the 
Milky Way with an N-body code (Dehnen et al., in preparation). 
These simulations suggest that only shortly after a tidal shock from a disk 
crossing the velocity dispersion in the cluster rises above the 
equilibrium dispersion whereas during other phases of the orbit 
the dispersion is settled down at the equilibrium level. 
Moreover, the numerical models confirm that at the cluster's present 
location near the apogalacticon of the orbit the line-of-sight velocity 
distribution should be nearly Gaussian. 
The observed departure from a Gaussian distribution must therefore indeed 
be due to binaries as we assumed above.  
 
The resulting very low dynamical velocity dispersion inside the cluster 
suggests that 
the velocities of the extratidal stars may locally also be very coherent.   
The narrow spatial confinement of the tidal tails, which is one of the 
reasons that led to their detection, is probably to some extent due to this 
kinematical coherence.   
We are currently conducting a similar kinematic study using the same 
instrumental equipment to investigate the velocities in the tidal tails 
of the cluster. Hereby we aim at measuring the velocity dispersion 
among the extratidal stars and the radial velocity gradient along the 
tails and hence along the orbit of the cluster.

\acknowledgments 
{\it Acknowledgements.} We thank ESO and the VLT staff for providing 
the observations in service mode. 

The Sloan Digital Sky Survey (SDSS) is a joint project of The University of 
Chicago, Fermilab, the Institute for Advanced Study, the Japan Participation 
Group, The Johns Hopkins University, the Max-Planck-Institute for Astronomy 
(MPIA), the Max-Planck-Institute for Astrophysics (MPA), New Mexico State 
University, Princeton University, the United States Naval Observatory, and 
the University of Washington. Apache Point Observatory, site of the SDSS 
telescopes, is operated by the Astrophysical Research Consortium (ARC). 
Funding for the project has been provided by the Alfred P. Sloan Foundation, 
the SDSS member institutions, the National Aeronautics and Space 
Administration, the National Science Foundation, the U.S. Department of 
Energy, the Japanese Monbukagakusho, and the Max Planck Society. 
The SDSS Web site is http://www.sdss.org/. 

%Funding for the creation and distribution of the SDSS Archive has been 
%provided by the Alfred P. Sloan Foundation, the Participating Institutions, 
%the National Aeronautics and Space Administration, the National Science 
%Foundation, the U.S. Department of Energy, the Japanese Monbukagakusho, 
%and the Max Planck Society. The SDSS Web site is http://www.sdss.org/. 
%The Participating Institutions are The University of Chicago, Fermilab, 
%the Institute for Advanced Study, the Japan Participation Group, The 
%Johns Hopkins University, Los Alamos National Laboratory, the 
%Max-Planck-Institute for Astronomy (MPIA), the Max-Planck-Institute for 
%Astrophysics (MPA), New Mexico State University, Princeton University, 
%the United States Naval Observatory, and the University of Washington.

The Digitized Sky Surveys were produced at the Space Telescope Science 
Institute under U.S. Government grant NAG W-2166. The images of these surveys 
are based on photographic data obtained using the Oschin Schmidt Telescope 
on Palomar Mountain and the UK Schmidt Telescope. 
The Second Palomar Observatory Sky Survey (POSS-II) was made by the 
California Institute of Technology with funds from the National Science 
Foundation, the National Geographic Society, the Sloan Foundation, 
the Samuel Oschin Foundation, and the Eastman Kodak Corporation. 
The Oschin Schmidt Telescope is operated by the California Institute of 
Technology and Palomar Observatory. \\

\begin{deluxetable}{ccccccc}
\tablecaption{Target identification and observing log\label{tab1}}
\tablewidth{0pt}
\tablehead{\colhead{Star} & \colhead{SH77} & \colhead{RA (2000)} & 
\colhead{DEC (2000)} & \colhead{ $r$ } & 
\colhead{Date of} & \colhead{Exp.\ Time} \\
& \colhead{ID} & \colhead{$[h\ m\ s]$} & \colhead{$[^{\circ}\ '\ '']$} & 
\colhead{ [$'$] } & \colhead{Observation} & 
\colhead{$[\ s\ ]$} }
\startdata
 1 & F  & 15 15 56.1 &$-$00 06 06 & 2.4 & 2001 May 04 &  420 \\ 
 2 & G  & 15 16 08.7 &$-$00 08 03 & 1.3 & 2001 May 04 &  460 \\ 
 3 & -- & 15 16 07.1 &$-$00 10 18 & 3.1 & 2001 May 04 &  510 \\ 
 4 & I  & 15 16 06.8 &$-$00 10 04 & 2.8 & 2001 May 04 &  660 \\ 
 5 & H  & 15 15 52.6 &$-$00 07 40 & 3.0 & 2001 May 04 &  660 \\ 
 6 & K  & 15 16 06.5 &$-$00 07 01 & 0.6 & 2001 May 10 &  730 \\ 
 7 & L  & 15 16 02.0 &$-$00 08 03 & 1.0 & 2001 May 10 &  810 \\ 
 8 & -- & 15 16 09.6 &$-$00 02 40 & 4.8 & 2001 May 10 &  810 \\ 
 9 & J  & 15 15 49.7 &$-$00 07 01 & 3.7 & 2001 May 10 &  880 \\ 
10 & N  & 15 15 59.5 &$-$00 08 59 & 2.1 & 2001 May 10 &  900 \\ 
11 & U  & 15 15 54.8 &$-$00 06 55 & 2.5 & 2001 May 10 & 1600 \\ 
12 & -- & 15 16 26.5 &$-$00 09 05 & 5.8 & 2001 May 10 & 1800 \\ 
13 & 6  & 15 15 58.3 &$-$00 09 46 & 3.0 & 2001 May 03 & 1800 \\ 
14 & 19 & 15 16 08.5 &$-$00 05 10 & 2.3 & 2001 May 04 & 2100 \\ 
15 & 27 & 15 16 00.3 &$-$00 06 00 & 1.6 & 2001 May 04 & 2400 \\ 
16 & AB & 15 15 48.2 &$-$00 06 07 & 4.2 & 2001 May 11 & 3600 \\ 
17 & 35 & 15 16 02.6 &$-$00 05 22 & 2.0 & 2001 Jun 23 & 2400 \\ 
18 & 26 & 15 16 04.8 &$-$00 06 28 & 0.8 & 2001 Jun 18 & 2700 \\ 
19 & -- & 15 16 20.7 &$-$00 07 33 & 4.0 & 2001 Jun 18 & 3000 \\ 
20 & 5  & 15 15 57.1 &$-$00 08 50 & 2.4 & 2001 Jun 19 & 3300 \\ 
HD\,107328&& 12 20 21.0& +03 18 45 & -- & 2001 May 04 &  1  \\
HD\,157457&& 17 26 00.0&$-$50 38 01& -- & 2001 May 04 &  10  \\
\enddata

Note:--SH77 = Sandage \& Hartwick (1977)\\
$r$ = angular distance from the center of Pal\,5 
\end{deluxetable}

\begin{deluxetable}{ccccccccc}
\tablecaption{Results from spectroscopy \label{tab2}}
\tablewidth{0pt}
\tablehead{
\colhead{Star} & Stellar &\colhead{MJD of}&\multicolumn{6}{c}{heliocentric 
radial velocity in \kms} \\ 
& type &\colhead{observation}&\colhead{$v_r$}&\colhead{$\epsilon$} 
& \colhead{$v_r$} & \colhead{$\epsilon$} & \colhead{$v_r$}&\colhead{$v_r$}\\
& & &  \multicolumn{2}{c}{(a)} & \multicolumn{2}{c}{(b)} & S85 & P85 }
\startdata
 1 & RGB & 52033.168  & $-58.51$ & 0.05& $-$58.66 & 0.05 & $-$54 & --  \\ 
 2 & RGB & 52033.178  & $-58.31$ & 0.05& $-$58.47 & 0.05 & $-$61 & --  \\ 
 3 & AGB  & 52033.187  & $-61.16$ & 0.06& $-$61.30 & 0.05 & --    & --  \\ 
 4 & MS & 52033.197  & $-23.44$ & 0.06& $-$23.62 & 0.06 & $-$23 & --  \\ 
 5 & AGB  & 52033.210  & $-56.92$ & 0.07& $-$57.06 & 0.05 & $-$53 & $-$46 \\ 
 6 & RGB & 52039.069  & $-58.72$ & 0.09& $-$58.87 & 0.11 & $-$61 & --  \\ 
 7 & RGB & 52039.082  & $-58.79$ & 0.10& $-$58.93 & 0.12 & $-$59 & --  \\ 
 8 & AGB  & 52039.096  & $-58.98$ & 0.05& $-$59.14 & 0.05 & --    & --  \\ 
 9 & AGB  & 52039.110  & $-57.35$ & 0.15& $-$57.49 & 0.15 & $-$57 & $-$55 \\ 
10 & RGB & 52039.124  & $-60.10$ & 0.14& $-$60.24 & 0.14 & $-$52 & --  \\ 
11 & RGB & 52039.138  & $-58.90$ & 0.09& $-$59.06 & 0.07 & --    & $-$58 \\ 
12 & MS & 52039.162  & $-52.97$ & 0.12& $-$53.13 & 0.12 & --    & --  \\ 
13 & RGB & 52032.268  & $-58.94$ & 0.08& $-$59.10 & 0.06 & --    & --  \\ 
14 & RGB & 52033.222  & $-61.07$ & 0.17& $-$61.19 & 0.19 & --    & --  \\ 
15 & RGB & 52033.251  & $-44.67$ & 0.12& $-$44.73 & 0.09 & --    & --  \\ 
16 & RGB & 52040.248  & $-58.33$ & 0.15& $-$58.50 & 0.16 & --    & --  \\ 
17 & RGB & 52083.033  & $-58.54$ & 0.06& $-$58.67 & 0.04 & --    & --  \\ 
18 & RGB & 52078.169  & $-58.42$ & 0.25& $-$58.55 & 0.22 & --    & --  \\ 
19 & RGB & 52078.130  & $-59.48$ & 0.22& $-$59.60 & 0.20 & --    & --  \\ 
20 & RGB & 52079.064  & $-57.27$ & 0.06& $-$57.40 & 0.05 & --    & --  \\ 
HD 107328 & & 52033.163  & --  &  -- &   +36.26 & 0.03 & & \\
HD 157457 & & 52032.438  &  +17.94  & 0.03&   --     &  --  & & \\
%HD 115404 & 52039.062  &   +6.00  & 0.19& +6.83    & 0.19 & MS& \\
\enddata

Notes:-- (a) by cross-correlation with HD\,107328\\
(b) by cross-correlation with HD\,157457 \\  
RGB = red giant branch, AGB = asymtotic giant branch, MS = main sequence 
star\\
$\epsilon$ = random error of the radial velocity measurement \\ 
S85 = Smith (1985), P85 = Peterson (1985)\\
\end{deluxetable}

\begin{deluxetable}{cccc}
\tablecaption{Membership probabilities from proper motion\label{tab3}}
\tablewidth{0pt}
\tablehead{\colhead{Star} & \colhead{SH77} & 
\colhead{CSM} & \colhead{ probability$^{(1)}$ } \\
& \colhead{ID} & \colhead{ ID } & \colhead{ in \% } }
\startdata
 1 & F  &     &  78 \\ 
 2 & G  &     &  73 \\ 
 3 & -- & 174 &  88 \\ 
 4 & I  &     &   0 \\ 
 5 & H  &     &  92 \\ 
 6 & K  &     &  88 \\ 
 7 & L  &     &  86 \\ 
 8 & -- &  32 &  72 \\ 
 9 & J  &     &  80 \\ 
10 & N  &     &  98 \\ 
11 & U  &     &  95 \\ 
12 & -- & 593 &   0 \\ 
13 & 6  &     &  98 \\ 
14 & 19 &     &  99 \\ 
15 & 27 &     &  73 \\ 
16 & AB &     &  80 \\ 
17 & 35 &     &  89 \\ 
18 & 26 &     &  97 \\ 
19 & -- & 142 &  99 \\ 
20 & 5  &     &  97 \\ 
\enddata

Notes:--SH77 = Sandage \& Hartwick (1977)\\
CSM = Cudworth, Schweitzer \& Majewski (in preparation) \\ 
$^{(1)}$\ according to the proper motion study by CSM \\
\end{deluxetable}

\begin{deluxetable}{clccccc}
\tablecaption{Maximum likelihood solutions for a Gaussian model plus binary component
\label{tab4}}
\tablewidth{0pt}
\tablehead{
\colhead{Model} & \colhead{Binary parameters} & \colhead{Sample} & \colhead{max($\ln \mathcal{L}$)} & 
\colhead{$x_{b}$} & \colhead{$\sigma$} & \colhead{$\sigma_{90\%}$} \\
& \colhead{and distributions}& & & & \kms\ & \kms\ } 
\startdata
--  &single stars only                 & R+A &$-$26.38& 0.0                    & $1.14^{+0.24}_{-0.19}$& 1.53 \\[2pt] 
(a) &standard binary model              & R+A &$-$25.94& $0.45^{+0.18}_{-0.26}$ & $0.24^{+0.57}_{-0.10}$& 1.07 \\[2pt] 
(b) &$e=0$ (*)                          & R+A &$-$26.08& $0.44^{+0.17}_{-0.27}$ & $0.24^{+0.59}_{-0.10}$& 1.08 \\[2pt] 
(c) &$f(e) = 2e$ (*)                    & R+A &$-$25.57& $0.50^{+0.21}_{-0.25}$ & $0.24^{+0.51}_{-0.11}$& 1.04 \\[2pt] 
(d) &$M_2 = M_1$ (*)                    & R+A &$-$26.38& $0.00^{+0.24}_{-0.00}$ & $1.06^{+0.38}_{-0.41}$& 1.39 \\[2pt] 
(e) &$M_2 = 0.4\,M_\odot$ (*)           & R+A &$-$25.98& $0.43^{+0.17}_{-0.26}$ & $0.24^{+0.56}_{-0.10}$& 1.07 \\[2pt] 
(f) &$M_2 = 0.2\,M_\odot$ (*)           & R+A &$-$25.16& $0.53^{+0.21}_{-0.23}$ & $0.24^{+0.45}_{-0.12}$& 1.02 \\[2pt] 
(g) &$M_2 = 0.2\,M_\odot, e = 0 $ (*)   & R+A &$-$25.32& $0.52^{+0.21}_{-0.23}$ & $0.24^{+0.47}_{-0.11}$& 1.03 \\[2pt] 
(h) &$M_2 = 0.2\,M_\odot, f(e)=2e$ (*)  & R+A &$-$25.02& $0.60^{+0.22}_{-0.23}$ & $0.23^{+0.46}_{-0.13}$& 1.01 \\
\\ 
\tableline
\\
--  &single stars only                  & R &$-$17.35& 0.0                    & $0.91^{+0.23}_{-0.18}$& 1.30 \\[2pt] 
(a) &standard binary model              & R &$-$15.30& $0.32^{+0.18}_{-0.16}$ & $0.23^{+0.20}_{-0.09}$& 0.74 \\[2pt] 
(b) &$e=0$ (*)                          & R &$-$15.38& $0.31^{+0.17}_{-0.16}$ & $0.23^{+0.21}_{-0.09}$& 0.75 \\[2pt] 
(c) &$f(e) = 2e$ (*)                    & R &$-$15.10& $0.35^{+0.19}_{-0.17}$ & $0.22^{+0.20}_{-0.09}$& 0.72 \\[2pt] 
(d) &$M_2 = M_1$ (*)                    & R &$-$16.54& $0.23^{+0.14}_{-0.15}$ & $0.24^{+0.35}_{-0.09}$& 0.89 \\[2pt] 
(e) &$M_2 = 0.4\,M_\odot$ (*)           & R &$-$15.37& $0.30^{+0.17}_{-0.15}$ & $0.23^{+0.20}_{-0.09}$& 0.75 \\[2pt] 
(f) &$M_2 = 0.2\,M_\odot$ (*)           & R &$-$14.78& $0.37^{+0.20}_{-0.17}$ & $0.22^{+0.19}_{-0.09}$& 0.69 \\[2pt] 
(g) &$M_2 = 0.2\,M_\odot, e = 0 $ (*)   & R &$-$14.87& $0.36^{+0.19}_{-0.17}$ & $0.22^{+0.19}_{-0.09}$& 0.70 \\[2pt] 
(h) &$M_2 = 0.2\,M_\odot, f(e)=2e$ (*)  & R &$-$14.67& $0.42^{+0.21}_{-0.18}$ & $0.22^{+0.19}_{-0.10}$& 0.69 \\[3pt] 
\enddata
\\
Notes:-- (*) = other parameters same as in model (a) \\
R+A = RGB and AGB stars ($n=17$), R = RGB stars only ($n=13$)\\
$x_{b}$ = best-fit binary fraction, $\sigma$ = Gaussian dispersion parameter\\
$\sigma_{90\%}$ = upper limit of $\sigma$ for 90\% confidence \\
%$q := |\ln \mathcal{L}_{obs} - <\ln \mathcal{L}>| / 
%\sqrt{<(\ln \mathcal{L} - <\ln \mathcal{L}>)^2>}$\\
 
\end{deluxetable}

%% Remarks for AAS manuscripts

%% Generally speaking, only the figure captions, and not the figures
%% themselves, are included in electronic manuscript submissions.
%% Use \figcaption to format your figure captions. They should begin on a
%% new page.

%% No more than seven \figcaption commands are allowed per page,
%% so if you have more than seven captions, insert a \clearpage

%% There must be a \figcaption command for each legend. Key the text of the
%% legend and the optional \label in curly braces. If you wish, you may
%% include the name of the corresponding figure file in square brackets.
%% The label is for identification purposes only. It will not insert the
%% figures themselves into the document.
%% If you want to include your art in the paper, use \plotone.
%% Refer to the on-line documentation for details.

%% Tables should be submitted one per page, so put a \clearpage before

%% Two options are available to the author for producing tables:  the
%% deluxetable environment provided by the AASTeX package or the LaTeX
%% table environment.  Use of deluxetable is preferred.
%%

%% Three table samples follow, two marked up in the deluxetable environment,
%% one marked up as a LaTeX table.

%% In this first example, note that the \tabletypesize{}
%% command has been used to reduce the font size of the table.
%% Note also that the \label command needs to be placed 

\clearpage

\begin{figure*}[t]
\includegraphics[scale=0.9,bb=100 280 435 565,clip=true]{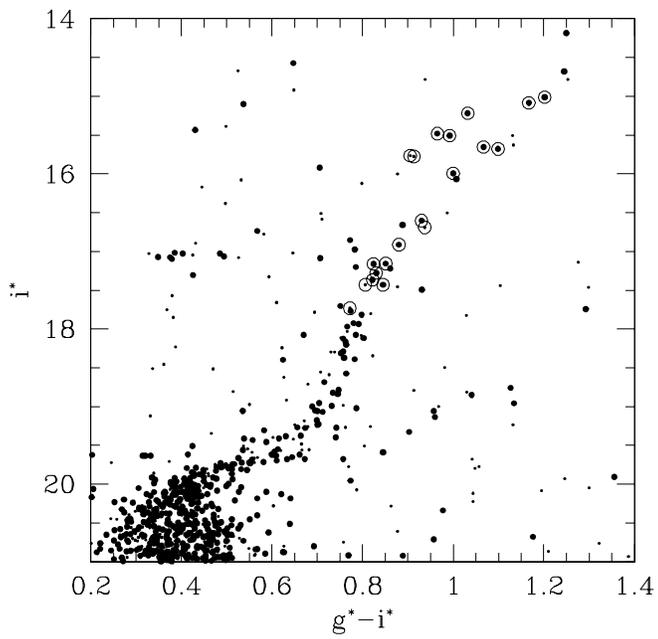}
\figcaption[fig1.eps]{Color-magnitude diagram of stars from the
Sloan Digital Sky Survey in the field of Pal\,5. Fat dots show stars 
with angular distance $r \le 3\farcm 6$ from the cluster center, 
small dots show stars with $3\farcm 6 < r \le 6\farcm 0$. 
The stars for which spectra were taken are marked by circles.  
\label{fig1}}
\end{figure*}

\clearpage

\begin{figure*}[t]
\includegraphics[scale=0.8,clip=true]{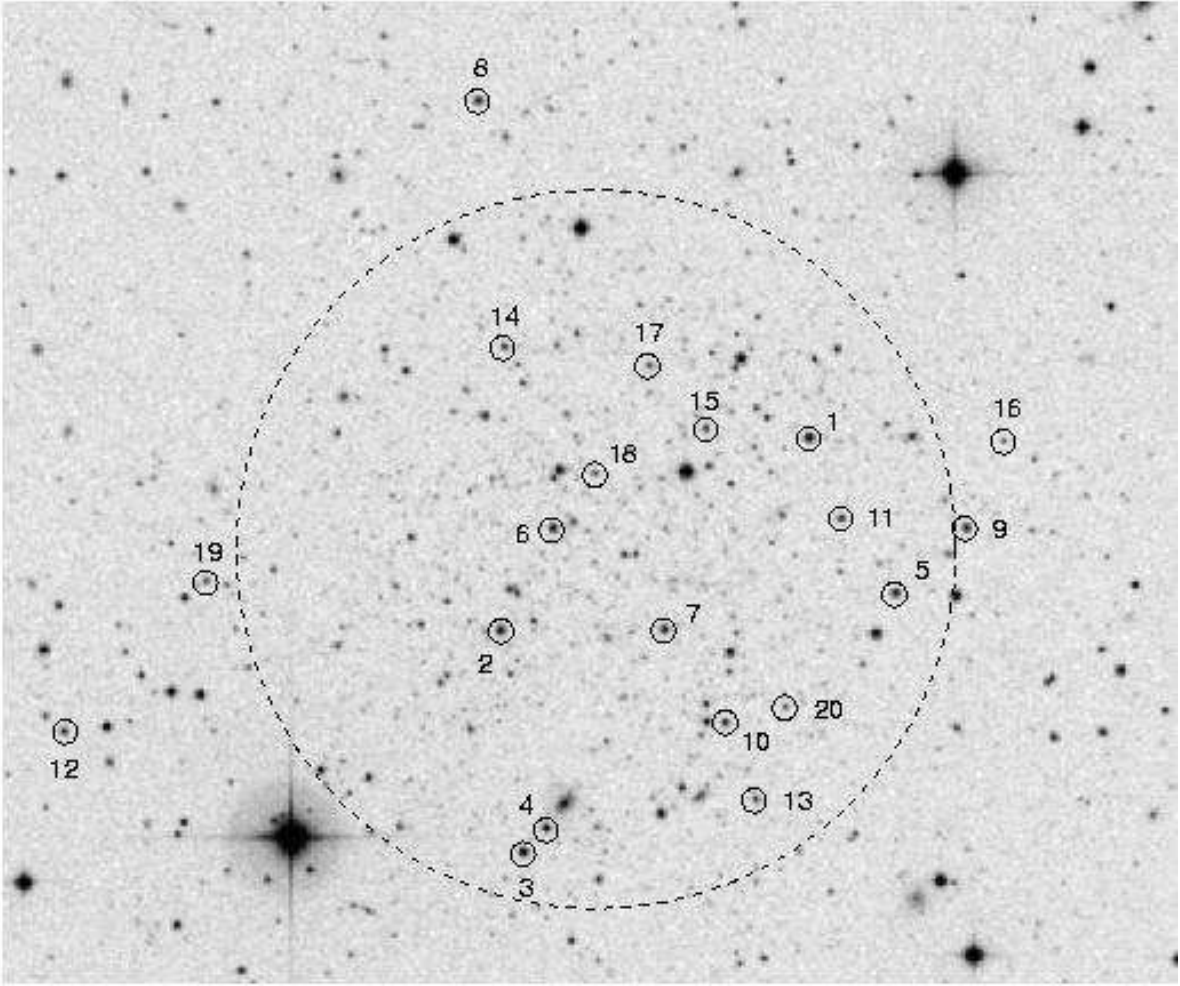}
\figcaption[fig2.eps]{Digitized Sky Survey image of Pal\,5 (DSS2) 
showing the positions of our spectroscopic targets 
(size $12' \times 10'$, north up, east to left). 
The region of the cluster core is marked by the dashed circle 
($r = 3\farcm 6$, core radius of the best-fitting King model, 
see Sect.\,5.1). 
\label{fig2}}
\end{figure*}

\clearpage

\begin{figure*}[t]
\includegraphics[scale=1.0,bb=100 280 510 600,clip=true]{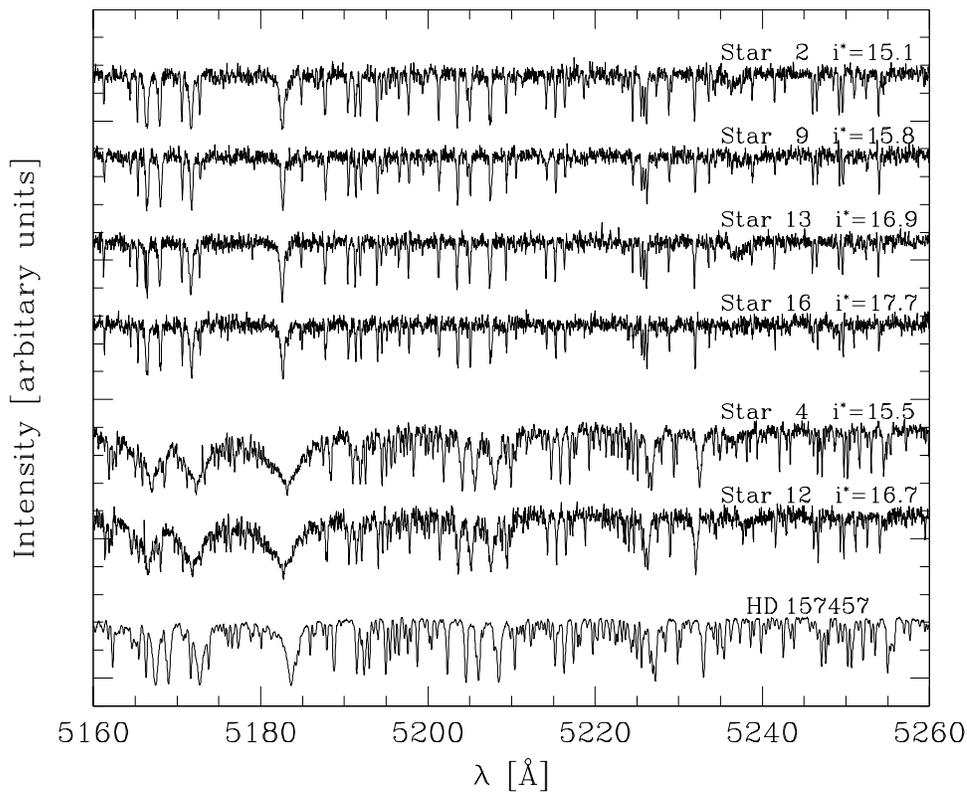}
\figcaption[fig3.eps]{Examples of the UVES spectra for program 
stars of different magnitude and for one of the standard stars. 
The plot shows a 100\,\AA\ wide section from the blue part of the 
spectrum, i.e., about 5\% of the full wavelength range covered by the 
spectra.
The left side of the plot contains the Mg~I~b triplet feature. Note 
the broad line profiles of stars 4 and 12, which indicate that they are 
dwarfs.    
\label{fig3}}
\end{figure*}

\clearpage

\begin{figure*}[t]
\includegraphics[scale=0.9,bb=100 175 435 700,clip=true]{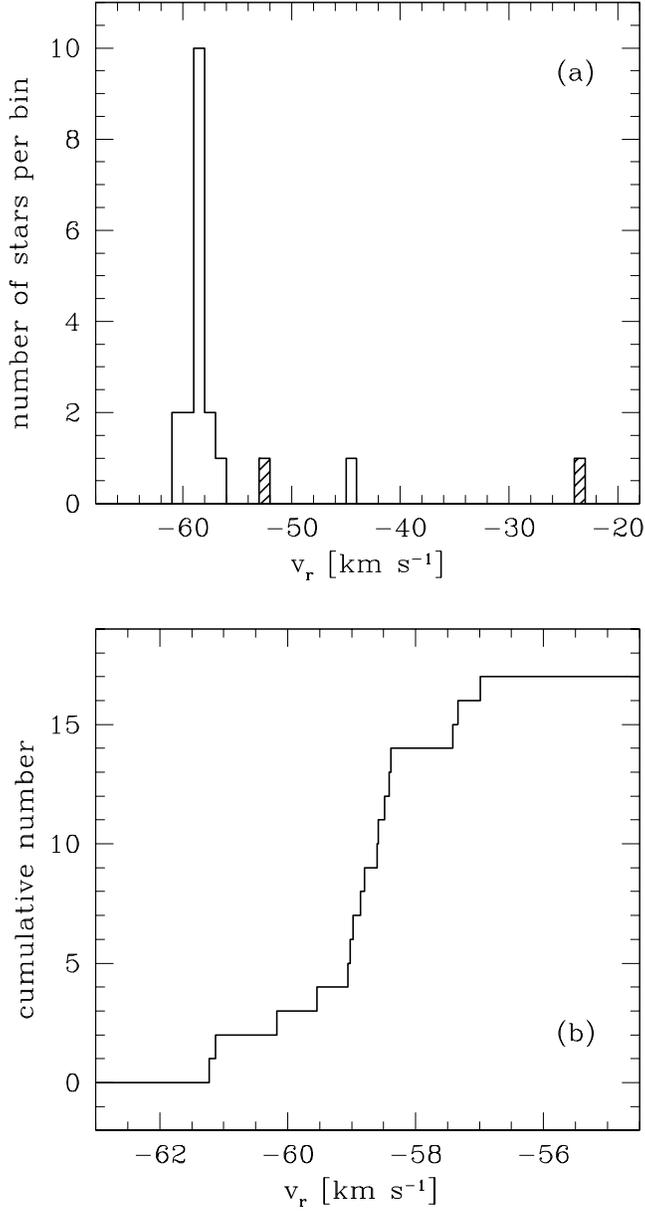}
\figcaption[fig4.eps]{Distribution of the heliocentric radial velocities
of the 20 program stars. 
(a) Histogram showing the number of stars in 1~\kms\ wide velocity bins. 
The shaded columns represent stars that have dwarf spectra and hence do 
not belong to Pal\,5. 
(b) Cumulative distribution, i.e., number of stars with velocities 
$\le v_r$, focussing on the range from $-$62~\kms\ to $-$56~\kms. 
\label{fig4}}
\end{figure*}

\clearpage

\begin{figure*}[t]
\includegraphics[scale=0.85,bb=18 370 570 630,clip=true]{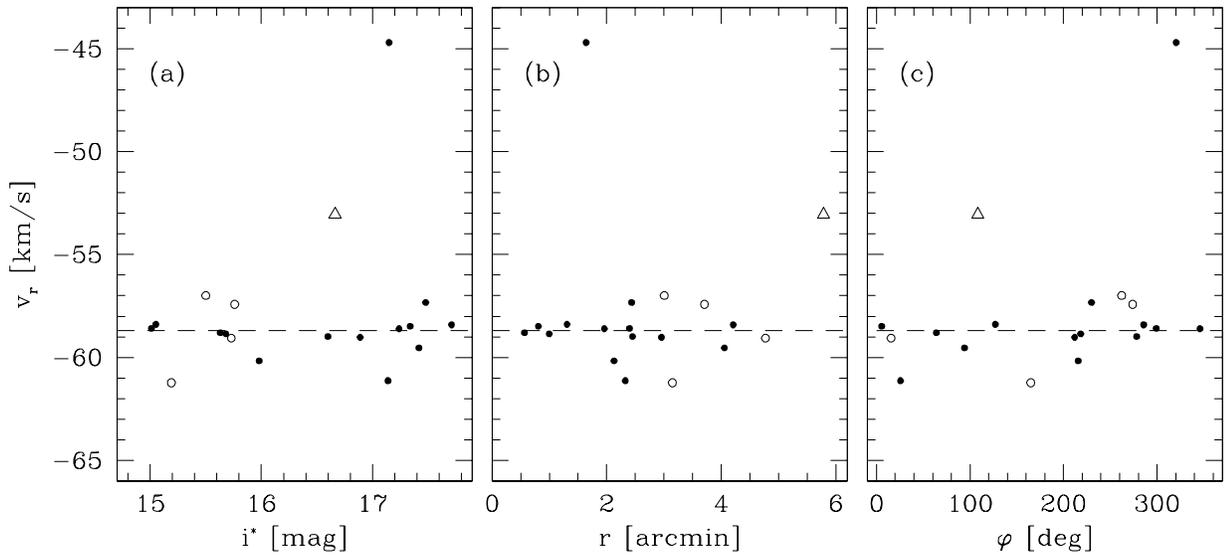}
\figcaption[fig5.eps]{Radial velocities of the program stars plotted 
versus brightness in $i^*$ (a), angular distance $r$ from the cluster center 
(b), and versus position angle $\varphi$ (c).
Filled circles show normal red giants while open circles show asymptotic 
giant branch stars. Open triangles mark stars with dwarf spectra.
The dashed line indicates the mean cluster velocity of $-$58.7~\kms. 
The measurement errors are in all cases smaller than the sizes of the plot 
symbols.   
\label{fig5}}
\end{figure*}

\clearpage

\begin{figure*}[t]
\includegraphics[scale=1.0,bb=100 200 475 670,clip=true]{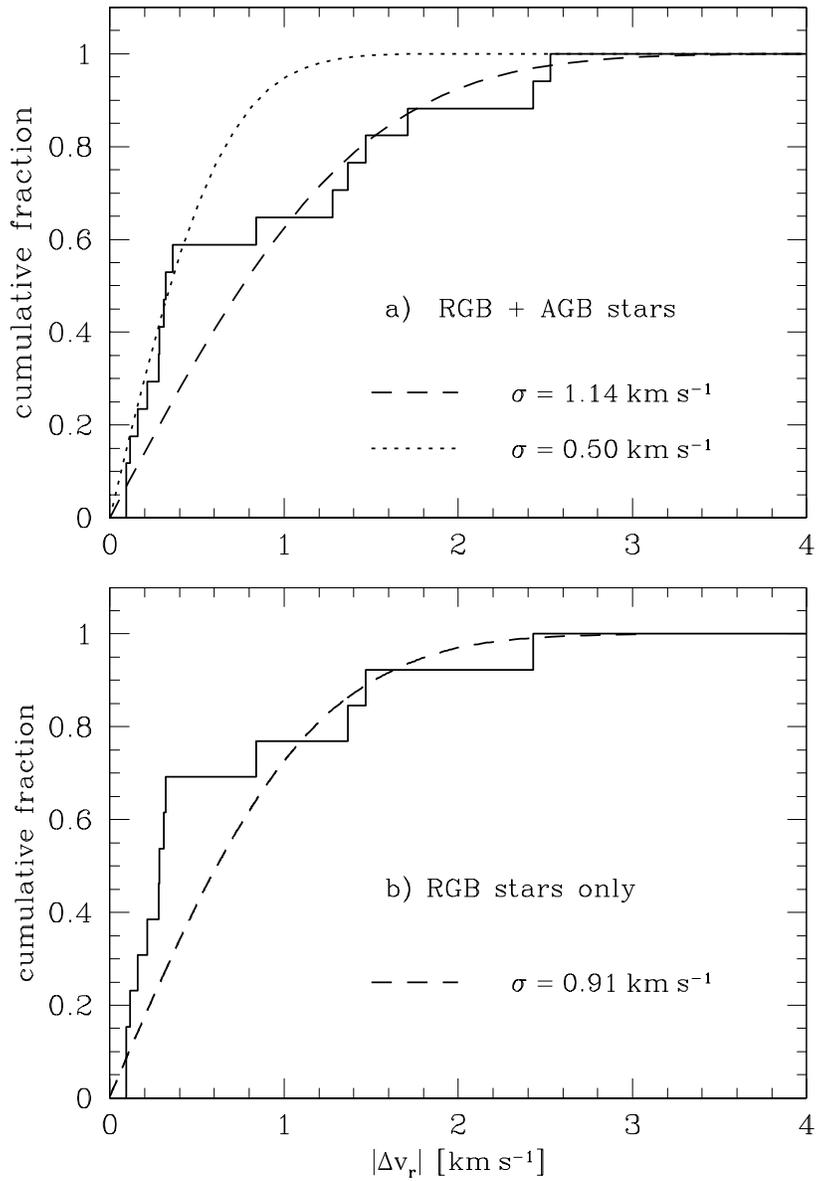}
\figcaption[fig6.eps]{
Cumulative distribution of the observed velocities (solid lines) 
compared with best-fit (error-convolved) Gaussian models (dashed lines).
a) Full sample of 17 certain cluster members and best-fit model 
with $\sigma = 1.14$~\kms. In addition, a Gaussian with $\sigma = 0.50$~\kms\ 
is shown (dotted line), which provides a better match to the inner
part of the observed distribution but fails in the outer part.  
b) Subsample of 13 RGB stars and best-fit model with $\sigma = 0.91$~\kms.
\label{fig6}}
\end{figure*}

\clearpage

\begin{figure*}[t]
\includegraphics[scale=1.0,bb=100 280 475 590,clip=true]{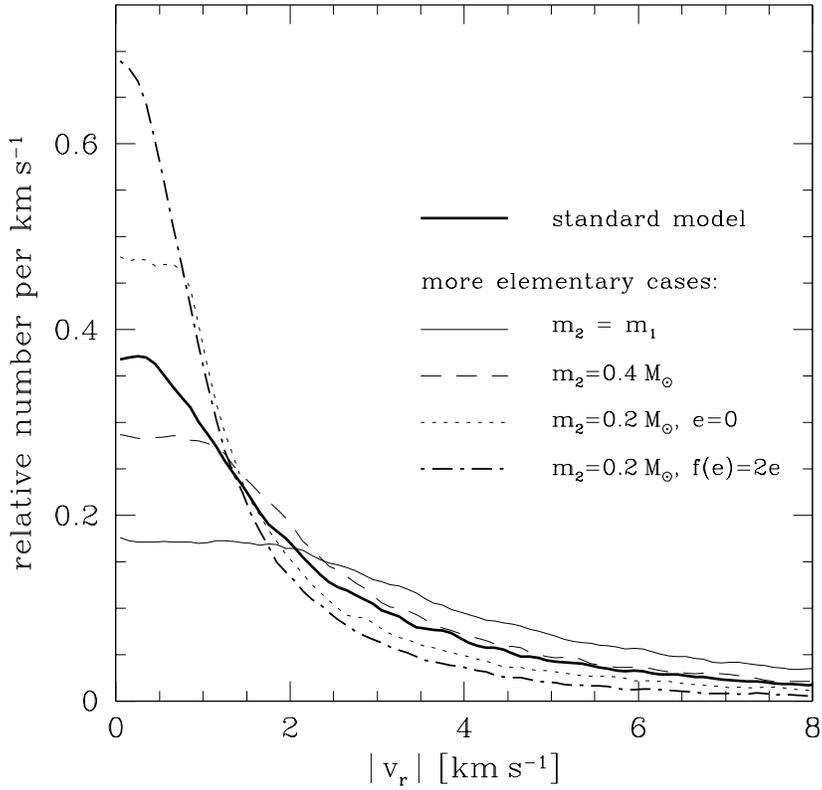}
\figcaption[fig7.eps]{Radial velocities induced by the orbital motion of 
a random-generated population of binaries with parameters as described 
in the text and in the legend. Here $v_r$ is the radial velocity 
of the primary component along isotropically distributed lines 
of sight at random orbital phase. 
The so-called standard binary model assumes a log-normal mass function 
for the secondary and a constant distribution of ellipticities as observed 
in local field binaries (for further details see Section 4). 
\label{fig7}}
\end{figure*}

\clearpage 

\begin{figure*}[t]
\includegraphics[scale=1.0,bb=90 265 465 590,clip=true]{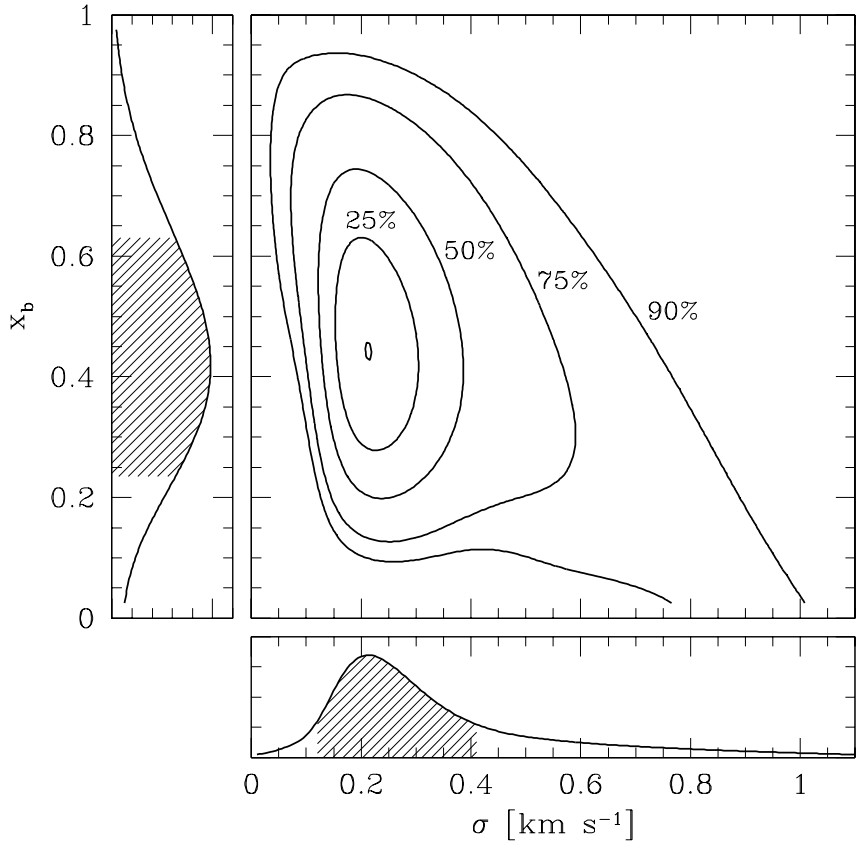}
\figcaption[fig8.eps]{Likelihood distribution in the plane of the 
parameters $x_b$ (binary frequency) and $\sigma$ (single star velocity 
dispersion) and marginal distributions thereof, calculated 
with the RGB star sample ($n=13$) and the binary model with $M_2=0.2~M_\odot$ 
and $f(e)=2e$ (see case (h) in the lower part of Tab.~4). 
The lines in the $x_b-\sigma$ plane show contours of equal likelihood 
$\mathcal{L}$ that encircle regions with 25\%, 50\%, 75\% and 90\% of the 
total integrated likelihood. The small central ellipse marks the location 
of the maximum of $\mathcal{L}$. Hatched areas mark the ranges of the 
probable values of $\sigma$ and $x_b$, i.e., the intervals around the 
maxima of $\sigma$ and $x_b$ that contain 68.3\% of the integrated 
likelihood.
\label{fig8}}
\end{figure*}

\clearpage

\begin{figure*}[t]
\includegraphics[scale=1.0,bb=100 200 475 670,clip=true]{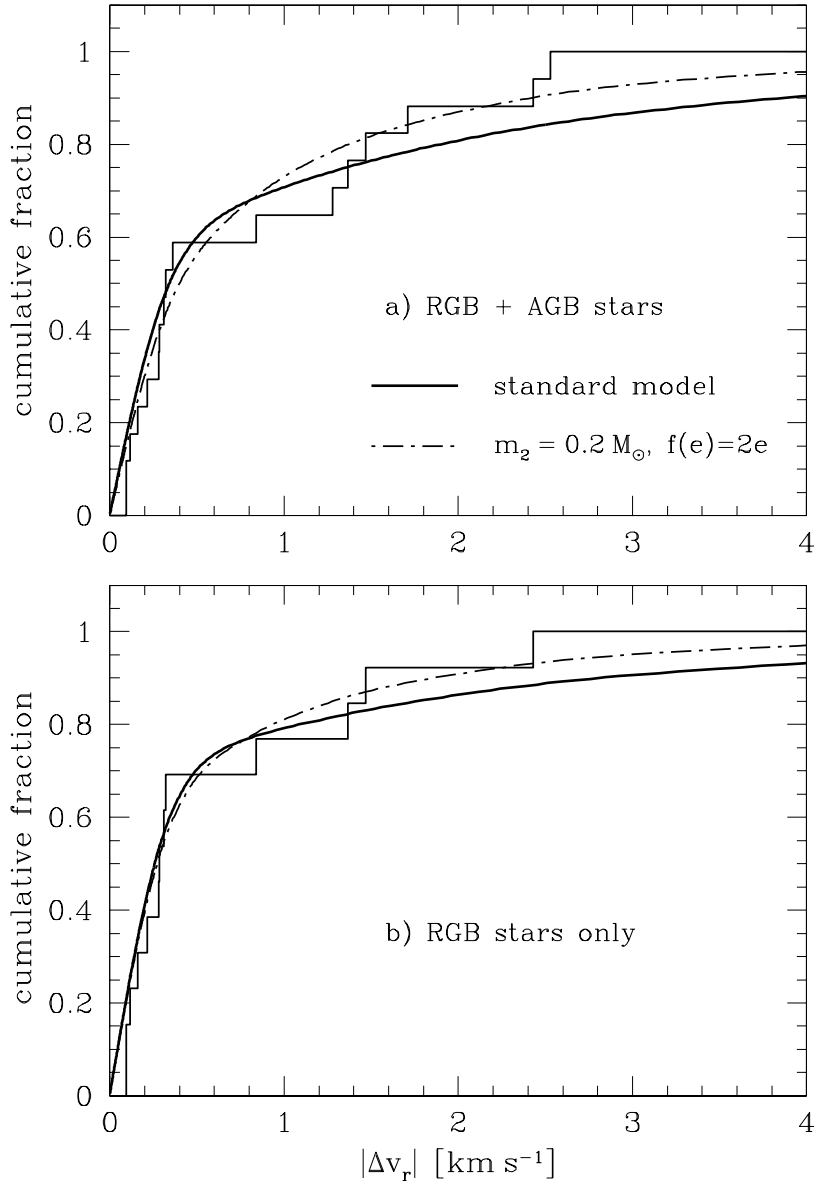}
\figcaption[fig9.eps]{Cumulative distribution of the observed velocities
(step functions, same as in Fig.~6) versus maximum likelihood predictions 
for an isothermal system with an adjustable fraction of binaries. 
For the parameters of the binary populations and the binary frequencies 
see Tab.~4. 
(a) Fit to all 17 cluster giants. (b) Fit to the subsample of 13 RGB stars.
\label{fig9}}
\end{figure*}

\clearpage

\begin{figure*}[t]
\includegraphics[scale=1.0,bb=100 280 465 590,clip=true]{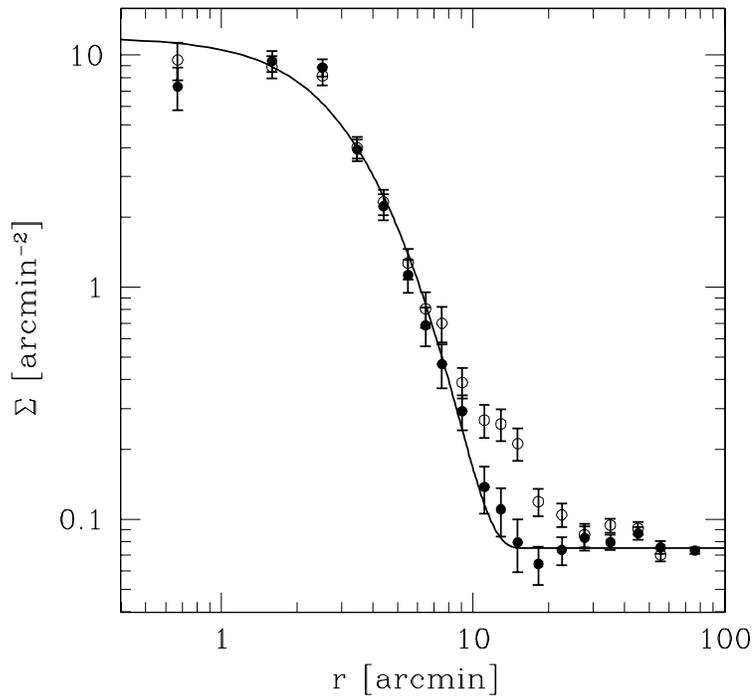}
\figcaption[fig10.eps]{Radial profile of the stellar surface density 
of Pal\,5 derived from SDSS. The datapoints with errorbars are the results 
of star counts in circular annuli around the cluster center in two 
complementary double sectors. 
Filled circles show the surface densities in the eastern and western sector 
($45^\circ \le PA \le 135^\circ$ and $225^\circ \le PA \le 315^\circ$)
while open circles are for the complementary northern and southern sector. 
The latter show an excess at $r > 6'$ that is due to extratidal stars. 
The solid line presents the best-fit King (1966) model (plus constant 
background) for the data shown by filled circles (parameters for that
model are $W_0 = 2.9$, $r_t = 16.1'$). 
\label{fig10}}
\end{figure*}

\clearpage

\end{document}